\documentclass[prl,superscriptaddress,twocolumn,floatfix]{revtex4}
\usepackage{amssymb}
\usepackage{amsmath}
\usepackage{graphicx}
\usepackage{color}

\newcommand{\VL}{$V_{\rm L}$}
\newcommand{\VR}{$V_{\rm R}$}

\newcommand{\gs}{$\rm{g_s}$}

\newcommand{\thirteen}{$^{13}$C}

\begin{document}
\title{Relaxation and Dephasing in a Two-electron $^{13}$C Nanotube Double Quantum Dot}
\author{H.~O.~H.~Churchill}
\affiliation{Department of Physics, Harvard University, Cambridge, Massachusetts 02138, USA}
\author{F.~Kuemmeth}
\affiliation{Department of Physics, Harvard University, Cambridge, Massachusetts 02138, USA}
\author{J.~W.~Harlow}
\affiliation{Department of Physics, Harvard University, Cambridge, Massachusetts 02138, USA}
\author{A.~J.~Bestwick}
\affiliation{Department of Physics, Harvard University, Cambridge, Massachusetts 02138, USA}
\author{E.~I.~Rashba}
\affiliation{Department of Physics, Harvard University, Cambridge, Massachusetts 02138, USA}
\affiliation{Center for Nanoscale Systems, Harvard University, Cambridge, Massachusetts 02138, USA}
\author{K.~Flensberg}
\affiliation{Nano-Science Center, Niels Bohr Institute, University of Copenhagen, Universitetsparken 5, DK-2100 Copenhagen, Denmark}
\author{C.~H.~Stwertka}
\affiliation{Department of Physics, Harvard University, Cambridge, Massachusetts 02138, USA}
\author{T.~Taychatanapat}
\affiliation{Department of Physics, Harvard University, Cambridge, Massachusetts 02138, USA}
\author{S.~K.~Watson}
\thanks{Present address:  Department of Physics, Middlebury College, Middlebury, Vermont 05753, USA}
\affiliation{Department of Physics, Harvard University, Cambridge, Massachusetts 02138, USA}
\author{C.~M.~Marcus}
\email{marcus@harvard.edu}
\affiliation{Department of Physics, Harvard University, Cambridge, Massachusetts 02138, USA}

\begin{abstract}
We use charge sensing of Pauli blockade (including spin and isospin) in a two-electron \thirteen~nanotube double quantum dot to measure relaxation and dephasing times.   The relaxation time, $T_1$, first decreases with parallel magnetic field then goes through a minimum in a field of 1.4 T.  We attribute both results to the spin-orbit-modified electronic spectrum of carbon nanotubes, which suppresses hyperfine mediated relaxation and enhances relaxation due to soft phonons.  The inhomogeneous dephasing time, $T_2^*$, is consistent with previous data on hyperfine coupling strength in \thirteen~nanotubes.
\end{abstract}

\maketitle
Few-electron double quantum dots have enabled the coherent manipulation and detection of individual and coupled electron spin states required to form qubits \cite{Loss-PRB98,Petta-Sci05,Koppens-Nat06,Hanson-RMP07}.
Although recent protocols mitigate decoherence due to hyperfine coupling in GaAs-based devices \cite{Koppens-PRL08,Reilly-Sci08}, an attractive alternative is to base spin qubits on group IV elements, which primarily comprise isotopes free of nuclear spins.
Progress in this direction includes double quantum dots in Si/SiGe 2DEGs \cite{Shaji-NPhys08}, P donors in Si \cite{Chan-JAP06}, Ge/Si nanowires \cite{Hu-NNano07}, and carbon nanotubes \cite{CNTDQDs}.
Recent advances in nanotube double dots include observation of singlet-triplet physics \cite{Jorgensen-NPhys08} and Pauli blockade \cite{Buitelaar-PRB08}.
Developing these systems as spin qubits depends crucially on understanding their modes of relaxation and dephasing.

%:Fig 1
%-------------------%
\begin{figure}[h!]
\center \label{figure1}
\includegraphics[width=3.2in]{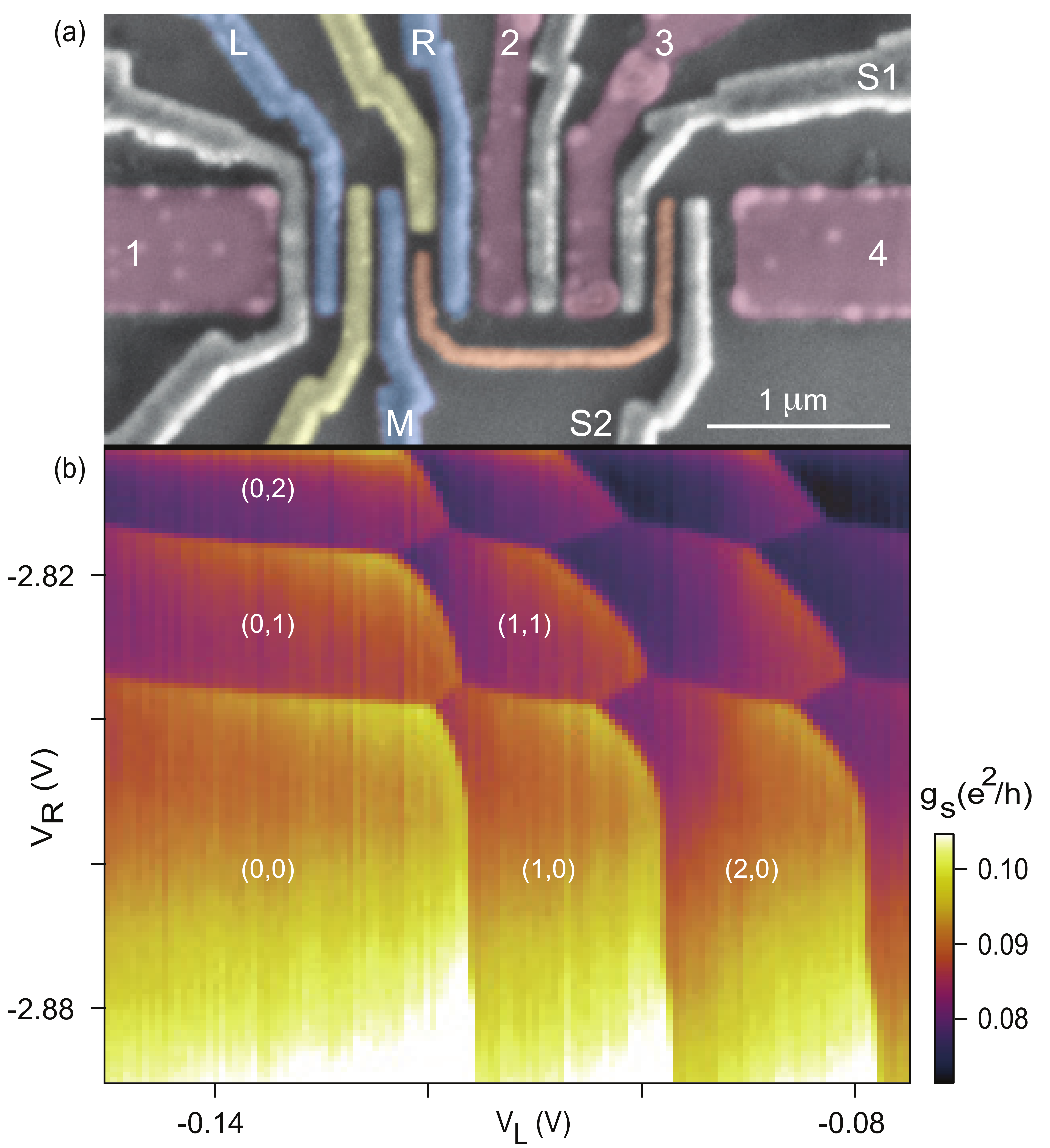}
\caption{\footnotesize{(a) False-color SEM micrograph of a device of the same design as the measured device.  The \thirteen\ nanotube (not visible) runs horizontally under Pd contacts (red).  The double dot is defined by top-gates L, R, and M (blue).  On the same nanotube, a separate quantum dot is controlled with gates S1 and S2 and capacitively coupled (orange wire) to the double dot to allow charge sensing.  Fast pulses are applied to L and R.  (b) Charge sensor conductance \gs~measured between contacts 3 and 4 as a function of \VL~and \VR~showing the charge stability diagram, with electron occupancies $(N_{\rm L},N_{\rm R})$ in each dot.}}
\end{figure}
%-------------------%

\par
This Letter reports measurements of relaxation and dephasing times in a two-electron nanotube double quantum dot grown from isotopically enriched (99\%) \thirteen~ methane. 
Measurements use fast pulses applied to electrostatic gates combined with charge sensing measurements in the Pauli blockade regime, including spin and isospin quantum states.
The relaxation time of these states, $T_1$,  initially decreases with parallel field and has a minimum in a field of 1.4 T. 
We interpret these results within the context of the recently observed \cite{Kuemmeth-Nat08} spin-orbit interaction in carbon nanotubes \cite{Ando-JPSJ2000,Bulaev-PRB08}. 
We also measure a relatively short two-electron inhomogeneous dephasing time, $T_2^*\sim 3$~ns, which presumably arises from hyperfine coupling. The implied hyperfine coupling strength is consistent with values measured recently by transport \cite{Paper1}. In contrast, the long $T_1$ does not show signatures of hyperfine coupling. 

\par
The double dot studied here is based on a single-walled carbon nanotube grown by chemical vapor deposition using 99\% \thirteen H$_4$ feedstock \cite{Liu-JACS01,methane}.
After deposition of two pairs of Pd contacts [Fig.~1(a), red], the device is coated with a 30 nm functionalized Al$_2$O$_3$ top-gate oxide using atomic layer deposition \cite{Jimmy-Sci07,Farmer-NL06}.
Aluminum top-gates (blue, yellow, and gray) define a double dot between contacts 1 and 2 and a single dot between contacts 3 and 4, capacitively coupled [orange wire in Fig.~1(a)] to the double dot to allow charge sensing \cite{Biercuk-PRB06,Hu-NNano07}.
The small bandgap ($\sim25$ meV) nanotube is operated in the electron regime.
Direct current and standard lock-in measurements are carried out in a dilution refrigerator (electron temperature $\sim100$ mK).

\par
Electron occupancies $(N_{\rm L},N_{\rm R})$ of the double dot are determined from the charge stability diagram (Fig.\ 1b), measured using the conductance, ${\rm g_s}$, of the charge-sensing dot \cite{Hu-NNano07}. 
Lever-arm ratios converting gate voltages to dot energies, extracted from nonlinear transport, give a large ($\sim1$ meV) interdot capacitive coupling, based on the size and shape of the stability diagram.
 
%:Fig 2
%-------------------%
\begin{figure}%[h!]
\center \label{figure2}
\includegraphics[width=3.4in]{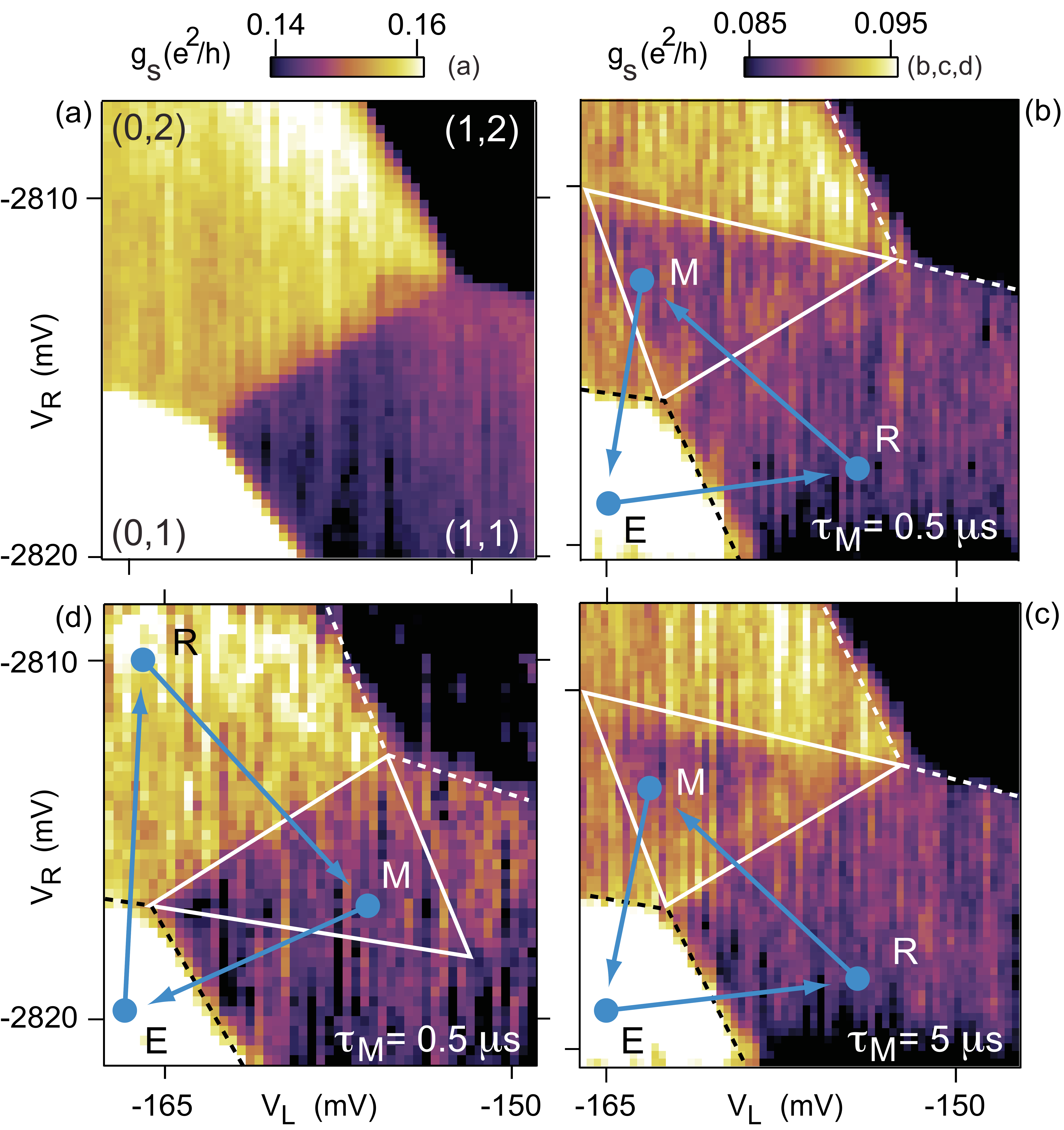}
\caption{\footnotesize{Sensor conductance \gs~as a function of \VR~and \VL~around the (1,1)/(0,2) transition (a) without applied pulses, (b) with the $T_1$ pulse cycle E$\rightarrow$R$\rightarrow$M$\rightarrow$E applied, ${\tau_{\rm M}}=0.5\ \mu$s.  
Dashed lines indicate the boundaries of (0,1) and (1,2) during step M.  
Within the pulse triangle (solid white lines), \gs~is between the (1,1) and (0,2) values, indicating partially blocked tunneling from (1,1) to (0,2), (c) with $T_1$ pulse cycle, ${\tau_{\rm M}}=5\ \mu$s, and (d) control pulse cycle, with R and M interchanged. $B=0$ in each panel.}}
\end{figure}
%-------------------%

\par
Single-electron states of a nanotube quantum dot (in the lowest circumferential mode) can be classified by a quantized longitudinal mode, a real spin ($S=1/2$), and an isospin, reflecting two valleys $K$ and $K'$ (or, equivalently, clockwise and counterclockwise motion around the nanotube circumference)~\cite{Kane-PRL97}.  
Including both spin and isospin, there are 16 ways to fill the lowest longitudinal modes with two electrons in the separated (1,1) charge state. There are only six ways, however, to fill the lowest longitudinal mode of (0,2) while maintaining overall antisymmetry of the wave function. 

\par

Under the condition of conserved spin and isospin in the double dot \cite{Pablo-Nat05}, the remaining 10 of the 16 two-electron states of (1,1) are blocked from tunneling to the lowest mode of (0,2) by selection rules on both spin and isospin. 
This is a generalization of the Pauli blockade \cite{Ono-Sci02,BriefTheory} observed in few-electron double dots without valley degeneracy. Previous experiments on Pauli blockade have only considered spin selection rules.

\par
Pauli blockade of the $(1,1)\rightarrow(0,2)$ transition is detected by time-averaged charge sensing, using the cyclic gate-pulse sequence in Fig.~2(b) \cite{Johnson-Nat05}:
Starting at E in (0,1), an electron is loaded with random spin and isospin, forming a (1,1) state at point R.
Moving to point M (adiabatically on the timescale of interdot tunnel coupling) where the ground state is (0,2) and remaining there for a time $\tau_{\rm M}$, the system may or may not tunnel to (0,2) depending on the state of (1,1). 
Blocked states would have to tunnel to higher-lying longitudinal modes of (0,2), which are energetically inaccessible at M ($\gtrsim1$ meV higher \cite{Paper1}); such states must flip either real spin or isospin (or both) to reach the lowest longitudinal mode.

\par
With the cycle E$\rightarrow$R$\rightarrow$M$\rightarrow$E running continuously, \VL~and \VR~are rastered in the vicintiy of the (1,1)-(0,2) charge transition [Fig.~2(b)]. 
Eighty percent of the pulse period is spent at M (10\% each for E and R) so that the time-averaged sensor signal ${\rm g_s}$ primarily reflects the charge state at M. 
Within the triangle marked by solid white lines in Figs.\ 2c-d, the time-averaged ${\rm g_s}$ lies between values on the (1,1) and (0,2) plateaus, decreasing in visibility as ${\tau_{\rm M}}$ is increased [Fig.~2(c)], with edges of the triangle disappearing faster due to thermal activation \cite{Johnson-Nat05}. We also observe faster relaxation within 200 $\mu$eV of the base. A control cycle with R and M interchanged does not show a triangular region in (1,1), indicating that none of the loaded $(0,2)$ states is blocked from tunneling into (1,1) [Fig.~2(d)].

%:Fig 3
%-------------------%
\begin{figure}%[h!]
\center \label{figure3}
\includegraphics[width=3.4in]{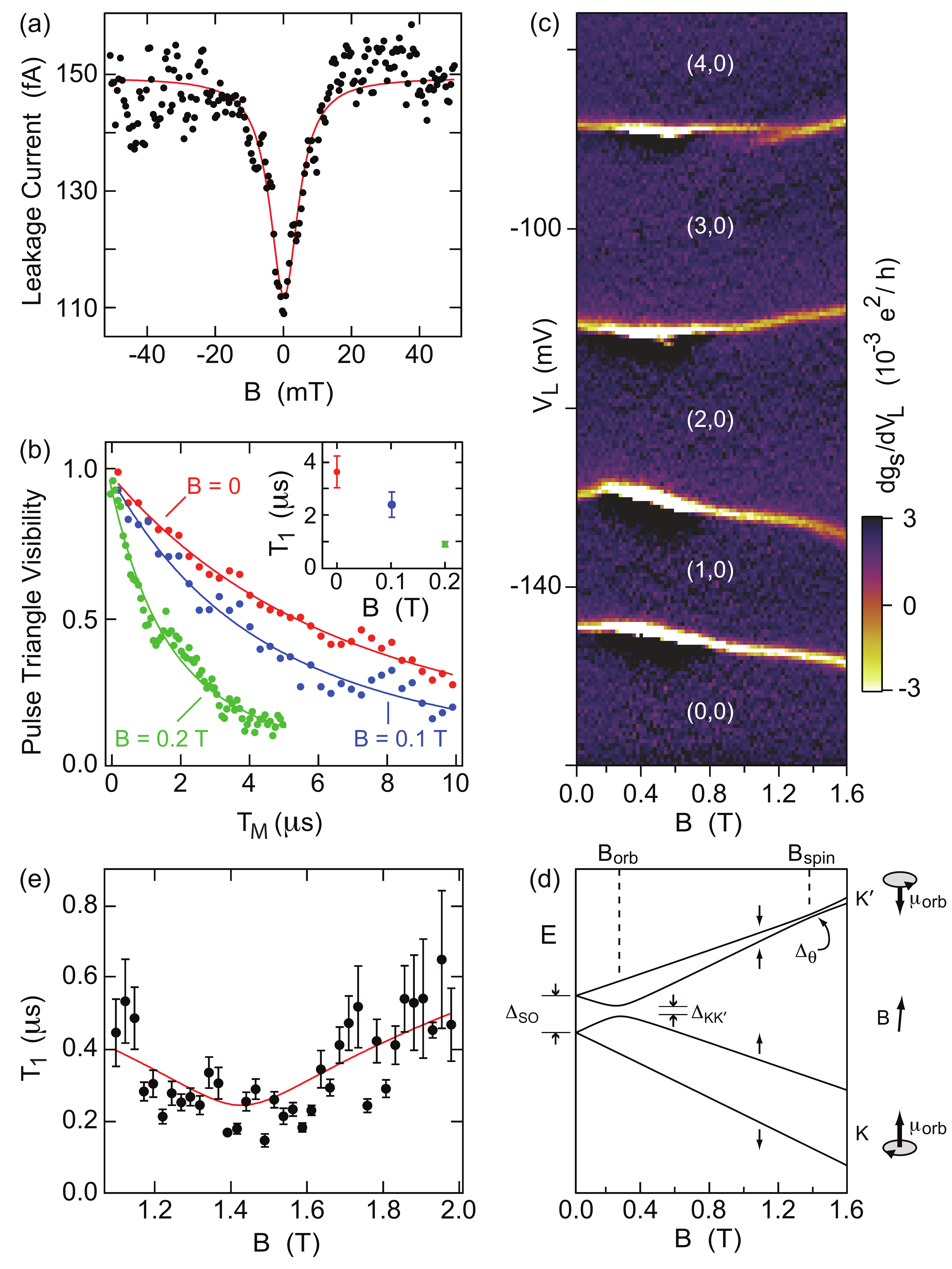}
\caption{\footnotesize{(a) Leakage current through blockade near zero detuning for small $B$, $V{\rm_{12}}=-2$ mV.  (b) Decay of pulse triangle visibility $I$ as a function of ${\tau_{\rm M}}$ measured in the center of the triangle at several values of $B$.  (c) ${\rm dg_s/d}V_L$ as a function of  \VL~and $B$, showing the dependence of ground state energies on $B$~for the first four electrons on the left dot.  (d) Energy level diagram of the lowest states of a nanotube with spin-orbit coupling; $\Delta_{\rm SO}=170\ \mu$eV, $\Delta_{\rm KK'}=25\ \mu$eV, $\theta=5^{\circ}$, and $\mu_{\rm orb}=330\ \mu$eV/T.  Arrows indicate spin component parallel to the nanotube axis.  Schematics (right) indicate orbital magnetic moment, $\mu_{\rm orb}$, for clockwise ($K$) and counterclockwise ($K'$) moving isospin states.  At $B_{\rm orb}$ ($B_{\rm spin}$), the orbital (Zeeman) shifts compensate $\Delta_{\rm SO}$ and states with opposite isospin (spin) anti-cross. (e) $T_1$ extracted  as in (b) for B~between 1.1 and 2 T.  Error bars: standard deviation of the fit parameter $T_1$.  One-parameter fit (red curve) to theory of Ref.~\onlinecite{Bulaev-PRB08}, modified for $B$~misaligned by 5$^{\circ}$ (see text).}}
\end{figure}
%-------------------%

\par
In a magnetic field, $B$, applied within a few degrees of parallel to the tube axis, forward bias ($V{\rm_{2}} >V{\rm_{1}}$) current---the Pauli-blockade direction---shows a dip around $B=0$ [Fig.~3(a)], indicating a reduced spin- and/or isospin-flip rate near zero field. In the reverse-bias case ($V{\rm_{1}} >V{\rm_{2}}$), current is independent of $B$ ($\sim1$ pA) over the same range.

\par
The pulse-triangle visibility, $I=\frac{{\rm g_s}(\tau_{\rm M})-{\rm g_s}(\infty)}{{\rm g_s}(0)-{\rm g_s}(\infty)}$ as a function of $\tau_{\rm M}$, measured in the center of the triangle [Figs.~2(b), (c)] at $B=0,\ 100,\ {\rm and}\ 200$ mT, is shown in Fig.\ 3(b) along with
the relaxation time $T_1$ extracted from fits to $I(\tau_{\rm M})=\frac{1}{\tau_{\rm M}}\int_0^{\tau_{\rm M}}e^{-t/T_1}\,dt$~\cite{Johnson-Nat05}. The relaxation time decreases with increasing $B$, but with a weaker dependence than the transport data  [Fig.~3(a)].
We speculate that these trends are due to phonon-mediated relaxation enabled by spin-orbit coupling \cite{Kuemmeth-Nat08,Bulaev-PRB08,Paper1Results}, a mechanism that is suppressed at small magnetic fields by Van Vleck cancellation \cite{Khaetskii-PRB01}. 

\par
Characteristics of the single-particle spectrum of the individual dots can be inferred from the $B$ dependence of the addition spectrum, measured for the left dot via charge sensing [Fig.\ 4(c)].  
Field dependences of the addition energies for the first four electrons suggest the spectrum shown in Fig.~4(d), consistent with Ref.~\onlinecite{Kuemmeth-Nat08} \cite{SOsign}, with spin-orbit coupling playing an important role.
We note, in particular, that the energy to add the second electron first increases with $B$ at small $B$, then decreases at higher field. This indicates that the second electron first occupies a counterclockwise ($K'$) isospin state at small $B$, then changes to a clockwise ($K$) isospin at $B\sim250$ mT. The energy to add the third electron does the opposite.
Fits to the low field slopes for the second and third electron addition energies yield moments of 390 $\mu$eV/T and $-270\ \mu$eV/T, respectively, with a difference in magnitudes within 10\% of 2$\mu_B$, a signature of a spin-orbit dominated spectrum \cite{Kuemmeth-Nat08}.
Thus we infer an orbital moment $\mu_{\rm orb}=330\ \mu$eV/T and a zero-field spin-orbit splitting $\Delta_{\rm SO}=170\ \mu$eV. 
Because the hyperfine coupling ($\sim 0.5~\mu$eV \cite{Paper1}) is much smaller than $\Delta_{\rm SO}$ and does not couple opposite isospins, relaxation between and within Kramer doublets due to random Overhauser fields from the \thirteen~nuclei is strongly suppressed.

\par
A consequence of the spectrum in Fig.\ 3(d) is a predicted \cite{Bulaev-PRB08} minimum in $T_1$ as the two $K'$ states with opposite spin approach one another at $B_{\rm spin}=\Delta_{\rm SO}/g\mu_B$, which for this nanotube occurs at 1.4~T [cf.~Fig.~3(d)]. 
The expected coupling of these two states is via 1D bending-mode phonons with quadratic dispersion, leading to a $T_1\propto\sqrt{\Delta}$ dependence on the energy splitting $\Delta$ due to the density-of-states singularity at zero energy in 1D \cite{Bulaev-PRB08}. This is in contrast  to higher dimensions, where  $T_1$ diverges as $\Delta \rightarrow 0$ \cite{Bulaev-PRB08, Khaetskii-PRB01, Amasha-PRL08}.

\par
Values for $T_1$, extracted from fits as in Fig.~3(b), are shown in Fig.~3(e), where a minimum in $T_1$ is observed at the predicted value, $B\sim1.4$ T.  
Also shown in Fig.~3(e) is a fit of the form $T_1=C\sqrt{\Delta_{\theta}}$, where the splitting $\Delta_{\theta}=g\mu_B\sqrt{(B\cos\theta-\Delta_{SO}/g\mu_B)^2+(B\sin\theta)^2}$ is anti-crossed, accounting for a misalignment angle, $\theta$, between the nanotube axis and the direction of the applied field~\cite{KKprime}. 
For these fits, we use $g=2$ and the measured quantities $\Delta_{SO}$ and $\theta$ ($5^{\circ}$ determined by the electron micrograph); the only free parameter is an overall scale for $T_1$,  $C=65$ ns/$\sqrt{\mu{\rm eV}}$, only a factor of $\sim5$ smaller than the estimates in Ref.~\onlinecite{Bulaev-PRB08}. 
Attributing the measured $T_1$ to this mechanism requires loading of at least one of the two higher states of Fig.~3(d) at step R, which is expected because the levels of the left dot are well below the electrochemical potential of the left lead at R.
We note that hyperfine relaxation should also be strongest near a degeneracy \cite{Johnson-Nat05}, but the ratio $\Delta_{\theta}/(g\mu_B B_{\rm nuc})\gg1$ \cite{Paper1} would require huge inelastic tunnel rates ruled out by transport measurements to explain the measured $T_1$.

\par 
While spin-orbit splitting suppresses relaxation via hyperfine interaction at $B = 0$, a difference in Overhauser fields between the two dots will induce dephasing of prepared two-particle spin and isospin states. 
To measure the inhomogeneous dephasing time $T_2^*$ of a state at $B=0$, a pulse cycle [Fig.~4(a)] first prepares an (0,2) state at P, then separates the electrons via P$'$ into (1,1) at S for a time $\tau_{\rm s}$, and finally measures the return probability to (0,2) at M \cite{Petta-Sci05}.
For small $\tau_{\rm s}$, the prepared state always returns to (0,2). For $\tau_{\rm s}\gtrsim T_2^*$, a fraction of prepared states evolves into blocked states, reducing the return probability within the pulse triangle [Fig.~4(a)].

%:Fig 4
%-------------------%
\begin{figure}%[h!]
\center \label{figure4}
\includegraphics[width=3.4in]{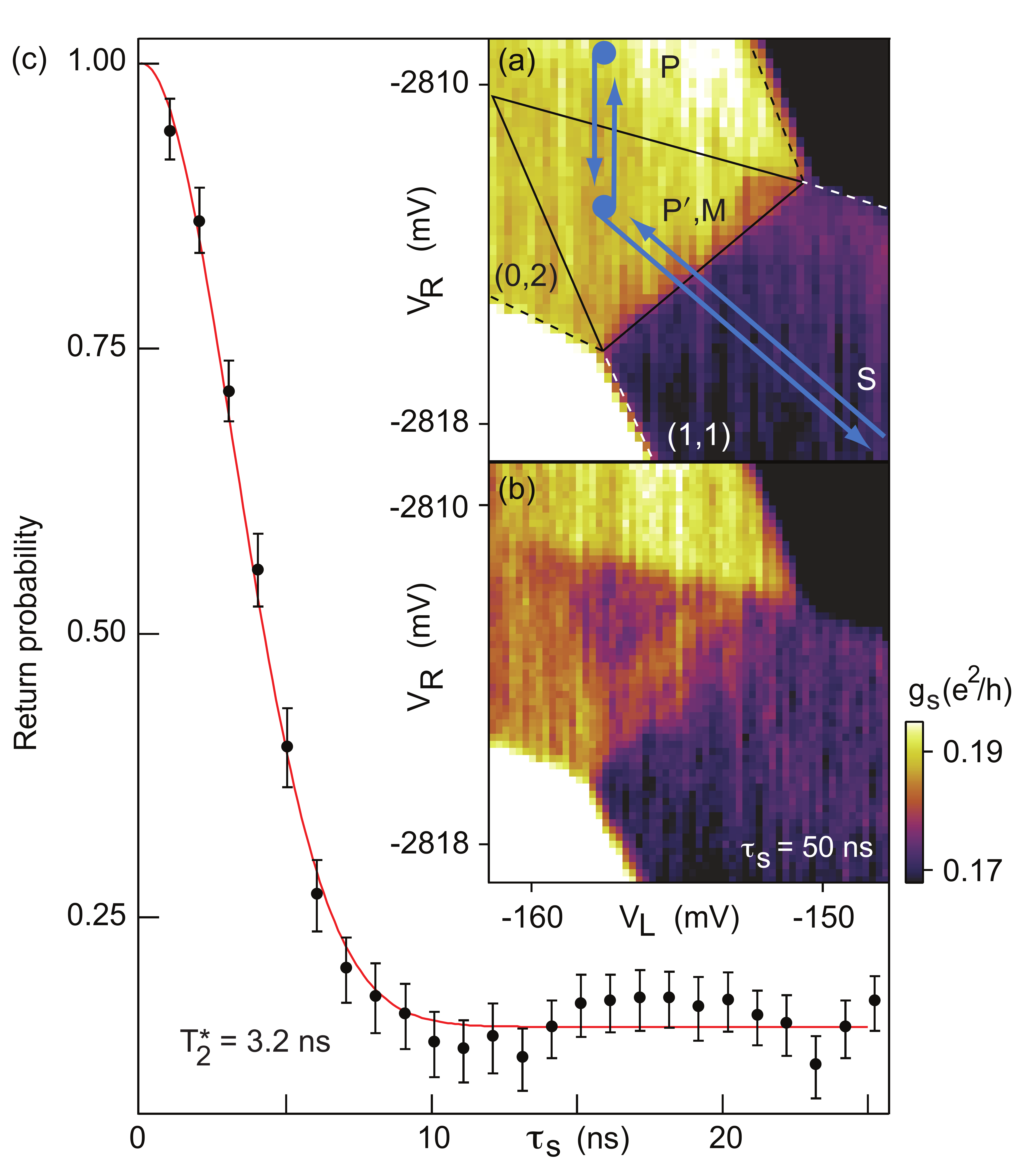}
\caption{\footnotesize{(a) Pulse sequence to measure the dephasing time $T_2^*$ (see text).  
If a state prepared at P dephases into a Pauli-blocked state while separated at S for a time $\tau_{\rm s}$, ${\rm g_s}$ is reduced within the pulse triangle outlined in black (shown in (b) for  $\tau_{\rm P}=\tau_{\rm P'}=100$ ns, $\tau_{\rm S}=50$ ns and $\tau_{\rm M}=2\ \mu$s at $B$$=0$).  (c) ${\rm g_s}$ calibrated to reflect the return probability to (0,2) versus $\tau_{\rm s}$.  A Gaussian fit (red) \cite{Taylor-PRB07} gives $T_2^*=3.2$ ns and ${\delta\rm B^{||}_{nuc}}=1.8$ mT. The data points are an average of 500 individual traces; error bars are the standard error.}}
\end{figure}
%-------------------%

\par
The dephasing time is obtained from the value of ${\rm g_s}$ in the center of the pulse triangle versus $\tau_{\rm s}$, which reflects the probability of return to (0,2) when calibrated against the equilibrium (1,1) and (0,2) values of ${\rm g_s}$ [Fig.~4(b)].
Assuming a difference in Overhauser fields acting on the two electrons of root mean square strength $\delta B_{\rm nuc}^{||}$ parallel to the nanotube axis \cite{Reilly-Sci08,Taylor-PRB07}, the decay is fit to a Gaussian form, giving $T_2^*=\hbar/g\mu_B\delta B_{\rm nuc}^{||}=3.2$ ns.
The corresponding $\delta B_{\rm nuc}^{||}=1.8$ mT is a factor of two smaller than our estimate of the single dot nuclear field $B_{\rm nuc}$ in \thirteen~nanotubes \cite{SecondDevice}.
The difference may be due to anisotropic dipolar hyperfine coupling \cite{Pennington-RMP96} or to accidental suppression of  $\delta B_{\rm nuc}^{||}$ \cite{Reilly-Sci08}.
The saturation value of the return probability in Fig.~4(c) is 0.17, smaller than the value of $1/3$ for singlet-triplet dephasing at $B=0$ in GaAs \cite{Petta-Sci05, Reilly08}, likely due to the richer spectrum allowed by isospin. 
Similarly, the tunneling probability from (1,1) to (0,2) (inferred from the visibility of the $T_1$ pulse triangle for $\tau_{\rm M} < T_1$, Fig.~2b) is lower than expected based on state-counting arguments (6 unblocked states out of 16 total) combined with adiabatic passage. This issue requires further study.

%:Summary
\par
In summary, we have measured relaxation and dephasing in a two-electron \thirteen~nanotube double quantum dot.
We identify signatures of spin-orbit coupling in the magnetic field dependence of both the addition spectrum and the relaxation time $T_1$, and we observed a dephasing time $T_2^*$ consistent with recent measurements of the hyperfine coupling strength in \thirteen~nanotubes.
The short dephasing time motivates development of nanotube devices with less than the 1\%~natural abundance of \thirteen.

%:Thanks
\par
We thank  C.~Barthel, E.~A.~Laird, and D.~J.~Reilly for technical assistance.  This work was supported by the National Science Foundation under grant no.~NIRT 0210736, ARO/iARPA, the Department of Defense, and Harvard's Center for Nanoscale Systems.  H.O.H.C. acknowledges support from the NSF.


\begin{thebibliography}{31}
\bibitem{methane}
\bibinfo{journal}{\thirteen H$_4$ of 99\% purity from Sigma-Aldrich, Co.  2 nm diameter is estimated from atomic-force microscope measurements of similar growths.}
\bibitem{Paper1Results}
\bibinfo{journal}{Results in Ref.~\onlinecite{Paper1} required many-electron configurations where these spin-orbit effects are absent or diminished.}
\bibitem{KKprime}
\bibinfo{journal}{This form is exact only for $\Delta_{KK'}=0$ but is justified because orbital mixing is suppressed by large $B$.}
\bibitem{SecondDevice}
\bibinfo{journal}{These data are for the device in \citep{Paper1} with $B_{\rm nuc}=4$ mT.}
\end{thebibliography}
\end{document}